 \documentclass{ws-ijmpe}

\usepackage[super,compress]{cite}
\usepackage{graphicx}

\usepackage{lineno}
\newcommand{\AuAu}{\ensuremath{{\rm Au+Au}}}
\newcommand{\pp}{$p$+$p$} 
\newcommand{\sqrtsNN}{\ensuremath{\sqrt{s_\mathrm{NN}}}}
\newcommand{\sqrts}{\ensuremath{\sqrt{s}}}

\newcommand{\pizero}{\ensuremath{\pi^0}}
\newcommand{\gammadir}{\ensuremath{\gamma_\mathrm{dir}}}

\newcommand{\pT}{\ensuremath{p_\mathrm{T}}}

\newcommand{\pTjet}{\ensuremath{p_\mathrm{T,jet}}}

\newcommand{\ETtrig}{\ensuremath{E_\mathrm{T}^\mathrm{trig}}}

\newcommand{\pTjetch}{\ensuremath{p_\mathrm{T}^\mathrm{jet,ch}}}

\newcommand{\gev}{\ensuremath{\mathrm{GeV/}c}}
\newcommand{\kT}{\ensuremath{k_\mathrm{T}}}

\newcommand{\rr}{\ensuremath{R}}

\newcommand{\RCP}{\ensuremath{R_\mathrm{CP}}}
\newcommand{\RAA}{\ensuremath{R_\mathrm{AA}}}
\newcommand{\RAAPythia}{\ensuremath{R^{\rm Pythia}_\mathrm{AA}}}

\newcommand{\IAA}{\ensuremath{{I}_{AA}}}
\newcommand{\IAApythia}{\ensuremath{I^\mathrm{PYTHIA}_{AA}}}

\newcommand{\Zg}{$z_{\mathrm g}$}
\newcommand{\Rg}{$R_{\mathrm g}$}
\begin{document}

\markboth{Nihar Sahoo}{An overview of recent STAR jet measurements}
%
\catchline{}{}{}{}{}
%

\title{An overview of recent STAR jet measurements}

\author{Nihar Ranjan Sahoo (for the STAR collaboration)}

\address{Institute of Frontier and Interdisciplinary Science, Shandong University, Qingdao, Shandong, 266237, China\\
Key Laboratory of Particle Physics and Particle Irradiation, Shandong University, Qingdao, Shandong, 266237, China
,\footnote{nihar@sdu.edu.cn, nihar@rcf.rhic.bnl.gov}\\
}

 
\maketitle

\begin{abstract}
These proceedings discuss recent jet measurements by the STAR experiment at RHIC to study jet substructure in \pp\ and jet quenching in Au+Au collisions at \sqrtsNN\ = 200 GeV. 
Furthermore, STAR's future plans for precision jet measurements with the upcoming data-taking periods in 2023-2025 are presented. 
\end{abstract}



\section{Introduction}
Jets in \pp\ and heavy-ion collisions arise from hard-scattered (high-$Q^{2}$) quarks and gluons of the incoming beams. In vacuum, a highly virtual parton generated in such interaction comes on-shell by radiating gluons, resulting in a jet shower. Studying jet properties in \pp\ collisions provides the opportunity to explore the perturbative and non-perturbative QCD effects in vacuum. In addition, the comparison between data and different QCD-based Monte Carlo (MC) event generators helps to constrain model parameters. In heavy-ion collisions, a highly energetic parton---while traversing through the Quark-Gluon Plasma (QGP)---interacts with the colored medium and loses its energy via medium-induced gluon radiation. This phenomenon is known as the jet quenching\cite{Cunqueiro:2021wls}. Suppression of jet yield, modification of jet shape and substructure, and jet deflection in the QGP are the manifestations of jet quenching in heavy-ion collisions. \\

 In these proceedings we present recent jet measurements by the STAR experiment at RHIC, addressing jet substructure in \pp\ collisions and jet quenching in Au+Au collisions.

\section{Jet measurements in \pp\ collisions}
\label{Sect:pp}
 The fragmentation and evolution of a hard-scattered parton is described by the Dokshitzer-Gribov-Lipatov-Altarelli-Parisi (DGLAP) splitting kernels~\cite{Dokshitzer:1977sg,Gribov:1972ri,Altarelli:1977zs}. This splitting probability depends on the momentum fraction of the split and the opening angle, and can be studied in \pp\ collisions. In QCD, higher order corrections contribute to the jet mass and substructure observables. To study these observables in STAR, jets are studied by clustering charged tracks from the time projection chamber and neutral particles from the barrel electromagnetic calorimeter towers using  the anti-\kT\ jet reconstruction algorithm~\cite{Salam:2010nqg} with different resolution parameters (\rr) between 0.2 and 0.6. Furthermore, such measurements in \pp\ collisions provide a baseline for the similar measurements in heavy-ion collisions to study the modification of parton shower in the finite-temperature QCD medium. 
\begin{figure}
    \centering
    \includegraphics[width=0.7\textwidth]{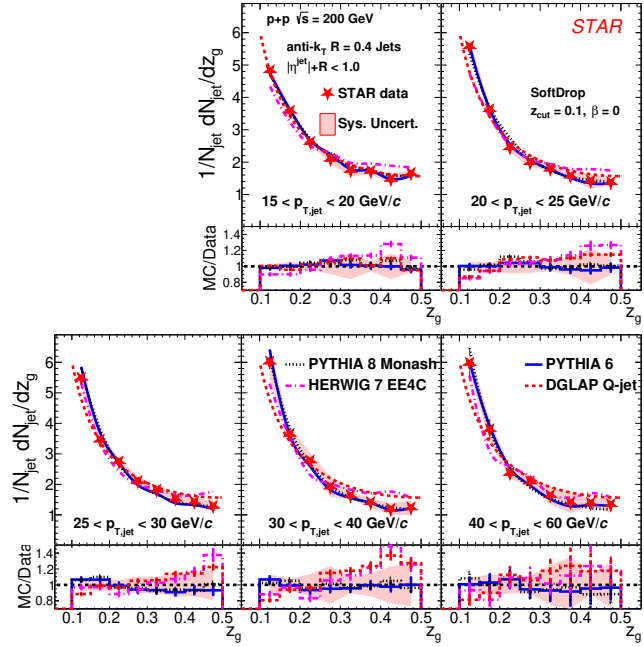}
    \caption{The \Zg\ distributions for different $p_{\rm T,jet}$ in \pp\ collisions at \sqrts\ = 200 GeV\cite{STAR:2020ejj}.}
    \label{fig:zg}
\end{figure}

\begin{figure}
    \centering
            \includegraphics[width=0.7\textwidth]{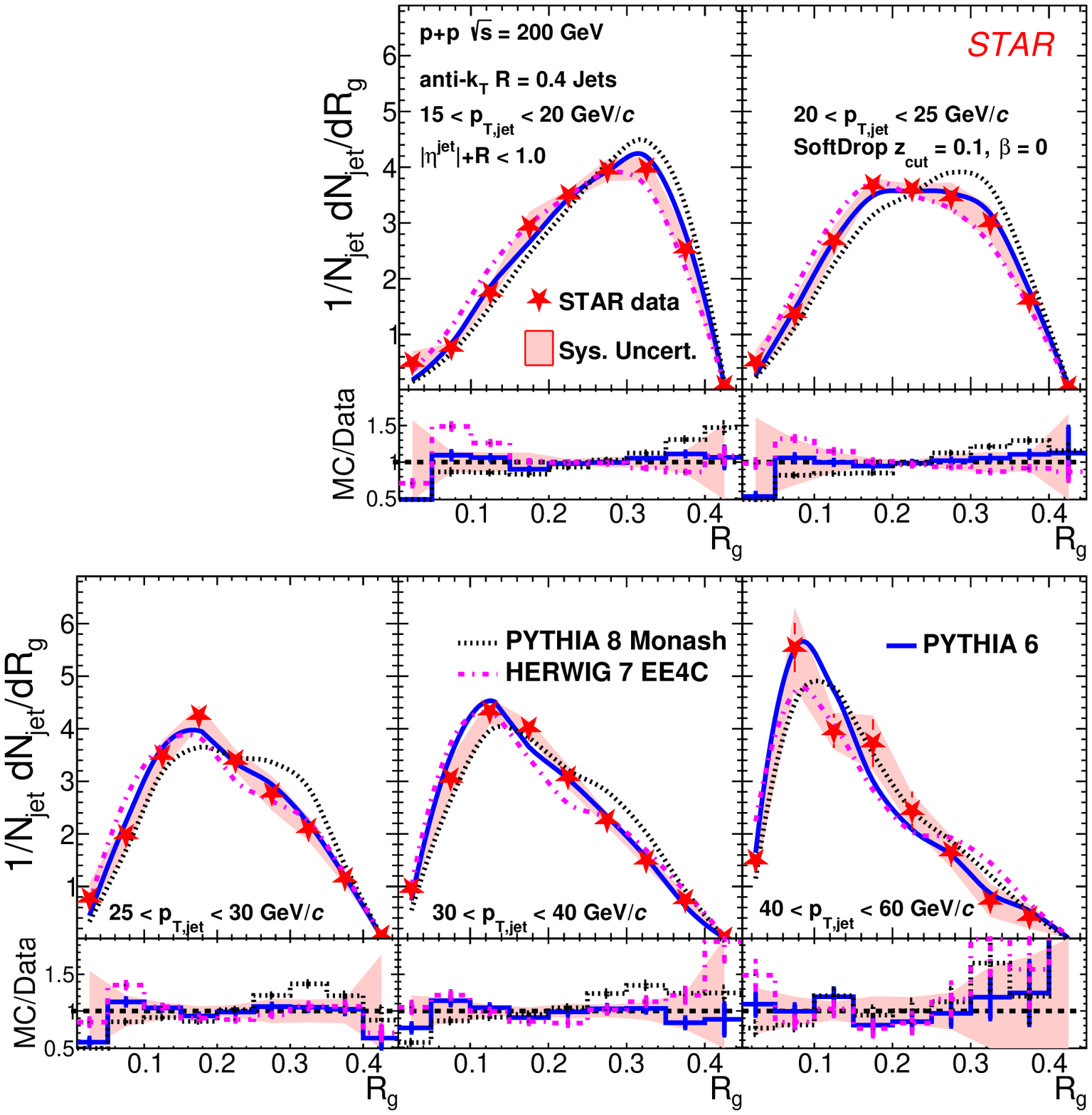}
    \caption{The \Rg\ distributions for different $p_{\rm T,jet}$ in \pp\ collisions at \sqrts\ = 200 GeV\cite{STAR:2020ejj}.}
    \label{fig:Rg}
\end{figure}

\begin{figure}
    \centering
    \includegraphics[width=0.7\textwidth]{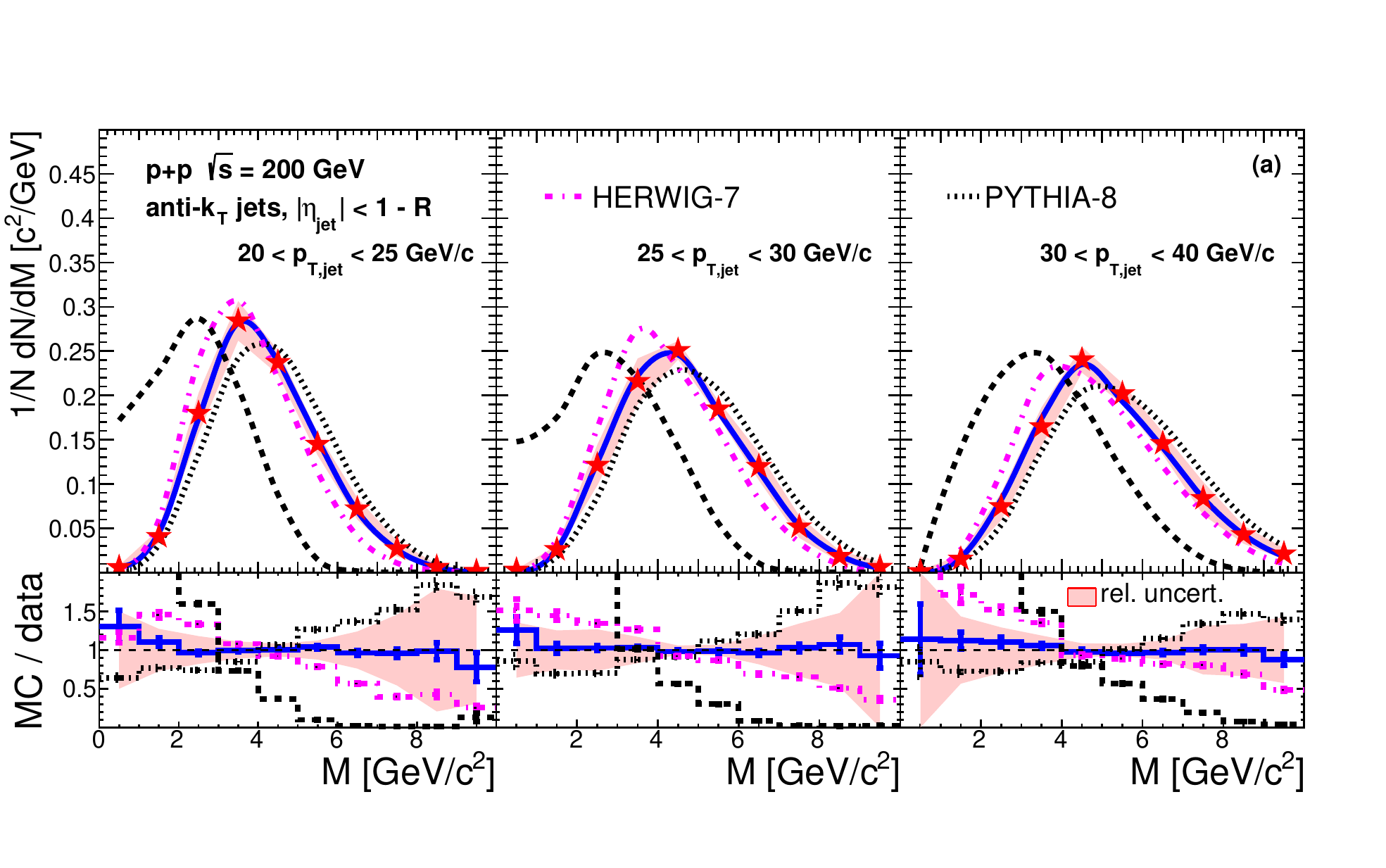}
        \includegraphics[width=0.7\textwidth]{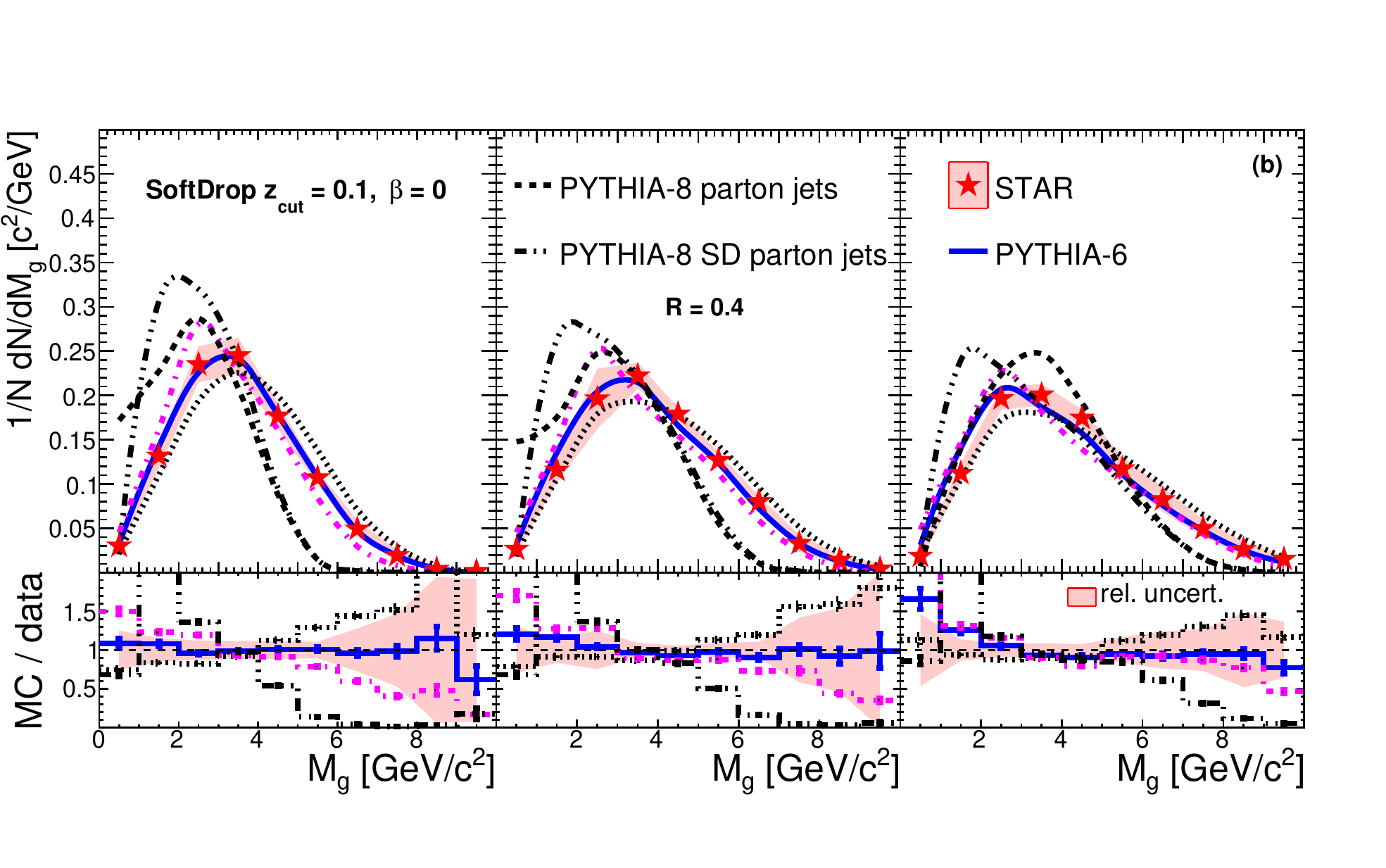}	
    \caption{Jet mass distributions for different $p_{\rm T,jet}$ and jet \rr\ in \pp\ collisions at \sqrts\ = 200 GeV\cite{STAR:2021lvw}.}
    \label{fig:Jetmass}
\end{figure}

\subsection{SoftDrop jet grooming}
The SoftDrop jet-grooming\cite{Larkoski:2015lea,Dasgupta:2013ihk,Larkoski:2014wba} algorithm helps to study jet substructure by suppressing soft large-angle radiations. In this procedure, the soft and wide-angle radiations are removed sequentially from the jet de-clustering tree. This is achieved using the Cambridge/Aachen (C/A) clustering algorithm\cite{Salam:2010nqg} by de-clustering jet branching history with removing the soft branch until it satisfies the condition: 

\begin{equation}
 z_{g} = \frac{ {\rm min} (p_{\rm T,1}, p_{\rm T,2})}{p_{\rm T,1}+p_{\rm T,2}} > z_{\rm cut} \big( \frac{R_{\rm g}}{R}\big)^{\beta}.   
\end{equation}
In Eq. (1), indices 1 and 2 represent the two sub-jets of splitting. The radius ($R$) is the distance in pseudorapidity and azimuthal angle space between two sub-jets, and \Rg\ is the groomed jet radius. The SoftDrop threshold $z_{\rm cut}=0.1$, and angular exponent $\beta=0$ are used for this de-clustering procedure for infrared and collinear (IRC) safety\cite{Larkoski:2014wba}.
\\
 
At RHIC, the first fully corrected SoftDrop observables, \Zg\ and \Rg , are measured in \pp\ collisions at $\sqrt{s}$ = 200 GeV by the STAR experiment for inclusive jets with \rr\ = 0.2, 0.4, and 0.6, and 15 $<$ \pTjet $<$ 60 \gev \cite{STAR:2020ejj}. Figures~\ref{fig:zg} and \ref{fig:Rg} show the distributions of \Zg\ and \Rg, respectively. The shape of \Zg\ distributions indicates no \pTjet\ dependence above 30 \gev, and they are more asymmetric than the DGLAP splitting function for a leading order quark emitting a gluon. \Rg\ distributions reveal a narrowing with increasing \pTjet, and the splitting is asymmetric at high \pTjet.  The STAR-tuned\cite{STAR:2019yqm} PYTHIA-6 with Perugia 2012 well describes the jet substructure observables at this energy. The comparisons with MC event generator predictions help further study different hadronization models for the higher-order QCD corrections at RHIC energy.

\subsection{Jet mass}
The mass of quark or gluon jets is sensitive to the fragmentation of highly virtual parent partons. The SoftDrop grooming procedure removes soft and wide-angle radiations from jets making the groomed jets less sensitive to the higher order QCD corrections. Jet mass is defined as the four-momentum sum of jet constituents, $M= \large| \sum_{i \in \rm jet}  p_{i}\large| =  \sqrt{E^{2}-{  {\textbf {\it p}}}^{2}}$. Here $E$ and ${\textbf {\it p}}$ are the energy and three-momentum of the jet, respectively.
 Studying both ungroomed and groomed jets, and comparing to different MC event generators can provide information on different pQCD effects and fragmentation. The STAR experiment has reported the first fully corrected ungroomed ($M$) and groomed ($M_{\rm g}$) mass distributions of inclusive jets for several values of \rr\ at $\sqrt{s}=200$ GeV as shown in Fig~\ref{fig:Jetmass}\cite{STAR:2021lvw}. These jets are selected within the range of $30 < \pTjet < 40 $ \gev. It is observed that the mean and width of the jet mass increases with increasing \rr\ due to the inclusion of wide-angle radiation. The same trend is also seen with growing \pTjet\ that increases the radiation phase space. The groomed jet mass distribution gets shifted to a smaller value than that of ungroomed mass due to the reduction of soft radiation in the SoftDrop grooming procedure. The LHC-tuned PYTHIA-8 and HERWIG-7 EE4C MC event generators over- and under-predicts the jet mass at RHIC, respectively, whereas the STAR-tuned PYTHIA-6 quantitatively describes the data. This observation is similar to that for the \Rg\ observable as discussed in the previous subsection. These measurements serve as a reference for future jet mass measurements in heavy-ion collisions at RHIC.

\section{Jet quenching measurements in Au+Au collisions}
\label{Sect:AuAu}
The STAR experiment has reported measurements of jet quenching using observable high-\pT\ hadron suppression\cite{STAR:2002ggv} and dihadron correlations\cite{STAR:2002svs,STAR:2006vcp}.The hadronic measurements have limited information on the underlying mechanism of jet quenching due to the final-state effects in heavy-ion collisions.  Over the last few years, the application of jet reconstruction algorithms and the development of methods for rigorous correction of uncorrelated background in heavy-ion collisions enable us to study jet quenching in more detail using fully reconstructed jets.\\
 
The first measurements of inclusive jet, semi-inclusive hadron+jet, and preliminary results of \gammadir+jet and \pizero+jet measurements have been reported by the STAR experiment. The measurement techniques and their results are discussed in this section.

\subsection{Inclusive jet suppression}
 Jet measurements in heavy-ion collisions are complicated due to the presence of large uncorrelated background. For the inclusive jet spectrum measurements in STAR, jets are reconstructed using anti-\kT\ algorithm\cite{Salam:2010nqg} with an additional requirement of a high-\pT\ hadronic constituent ($p_{\rm T,lead}^{\rm min}$), in order to identify jets from hard scattering processes. The selection of $p_{\rm T,lead}^{\rm min}$ should satisfy the following criteria: i) it must be sufficiently high so that contributions from combinatorial jets are negligible; ii) the probability of multiple constituents with \pT\ $\geq p_{\rm T,lead}^{\rm min}$ is negligible; iii) this $p_{\rm T,lead}^{\rm min}$ cut does not introduce a selection bias on the jet population within the considered $p_{\rm T,jet}$ range. Using this technique, the first fully corrected inclusive jet spectrea in central and peripheral Au+Au collisions at \sqrtsNN\ = 200 GeV have been reported with $p_{\rm T,lead}^{\rm min}$=5 \gev\ \cite{STAR:2020xiv}. \\
 
\begin{figure}
    \centering
    \includegraphics[width=0.9\textwidth]{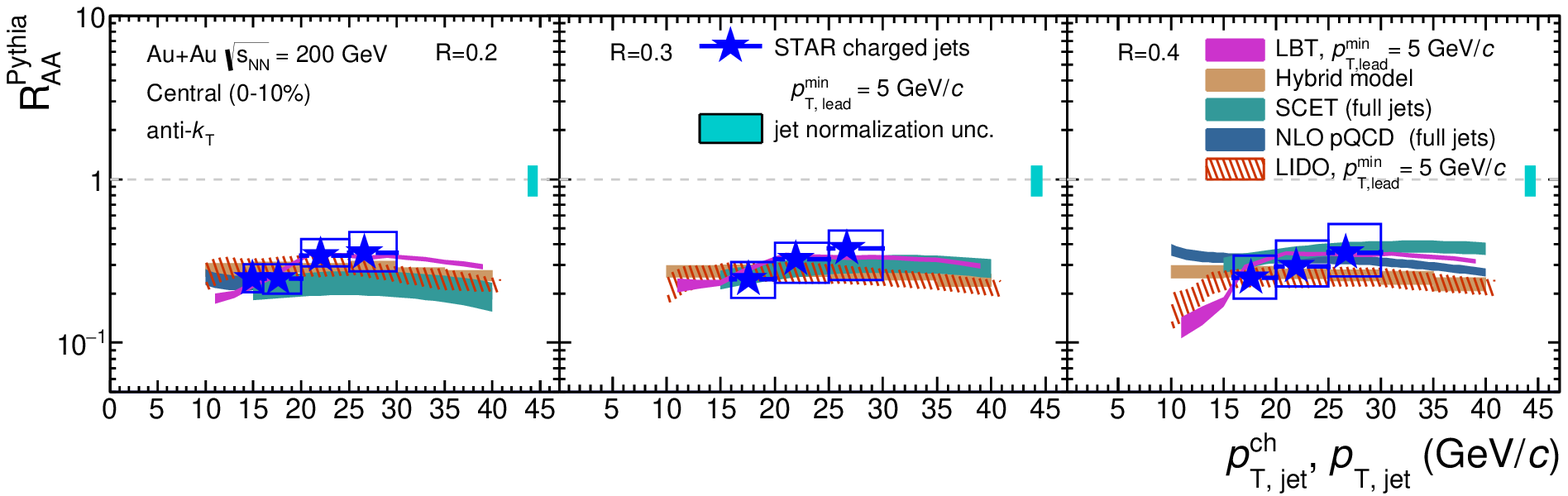}
    \caption{\RAAPythia\ as a function of \pTjet\ in 0-10\% central Au+Au collisions and for different jet \rr\ at \sqrtsNN\ = 200 GeV\cite{STAR:2020xiv}.}
    \label{fig:InclJetRAa}
\end{figure}

\begin{figure}
    \centering
    \includegraphics[width=0.9\textwidth]{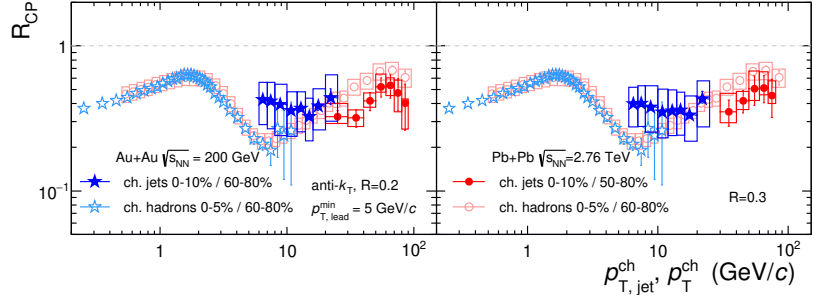}
    \caption{\RCP\ as a function of \pTjet\ in Au+Au collisions and for different jet \rr\  at \sqrtsNN\ = 200 GeV\cite{STAR:2020xiv}.}
    \label{fig:InclJetRCP}
\end{figure}

 The nuclear modification factor (\RAA) is defined as the ratio of inclusive charged jet yield in central Au+Au collisions to its cross sections in \pp\ collisions scaled by the nuclear thickness factor $\langle T_{\mathrm{AA}} \rangle$ of central collisions. Similarly, \RCP\ is defined considering 60-80\% peripheral collisions as a reference instead of \pp\ collisions. For \RAA,  PYTHIA is used as a vacuum reference, hence it is labeled as \RAAPythia. Figure~\ref{fig:InclJetRAa} shows the \RAAPythia\ as a function of $p_{\rm T,jet}^{\rm ch}$ for inclusive jets with $\rr=0.2, 0.3$ and 0.4 within 15 $<p_{\rm T,jet}^{\rm ch}<$ 30 \gev. Strong suppression is observed in 0-10\% central \AuAu\ collisions, and no jet \rr\ dependence of the suppression is seen. Different theory calculations\cite{Vitev:2009rd,Chien:2015hda,Casalderrey-Solana:2016jvj,He:2015pra,Ke:2018jem}are consistent with the data. The \RCP\ shows strong and similar suppressions for $\rr=0.2$ and 0.3 jets as shown in Fig~\ref{fig:InclJetRCP}. The yields of inclusive charged hadrons and jets show a comparable level of suppression within the same \pT\ interval in central Au+Au collisions at \sqrtsNN\ = 200 GeV. In addition, the comparison between central Au+Au collisions at RHIC and central Pb+Pb collisions at the LHC---although within different \pT\ intervals---show a similar magnitude of suppression for charged hadrons and jets yields.
 The medium-induced broadening of the inclusive jet is measured by taking the ratio of inclusive jet yields for $\rr=0.2$ and 0.4. No significant modification of the transverse jet profile due to jet quenching is observed for the inclusive jet population in central Au+Au collisions at $\sqrtsNN\ =  200$ GeV and is consistent with the LHC data. The ongoing full jet analysis will access the inclusive jet suppression measurement at higher \pTjet\ at RHIC energies.

\begin{figure}
    \centering
    \includegraphics[width=1\textwidth]{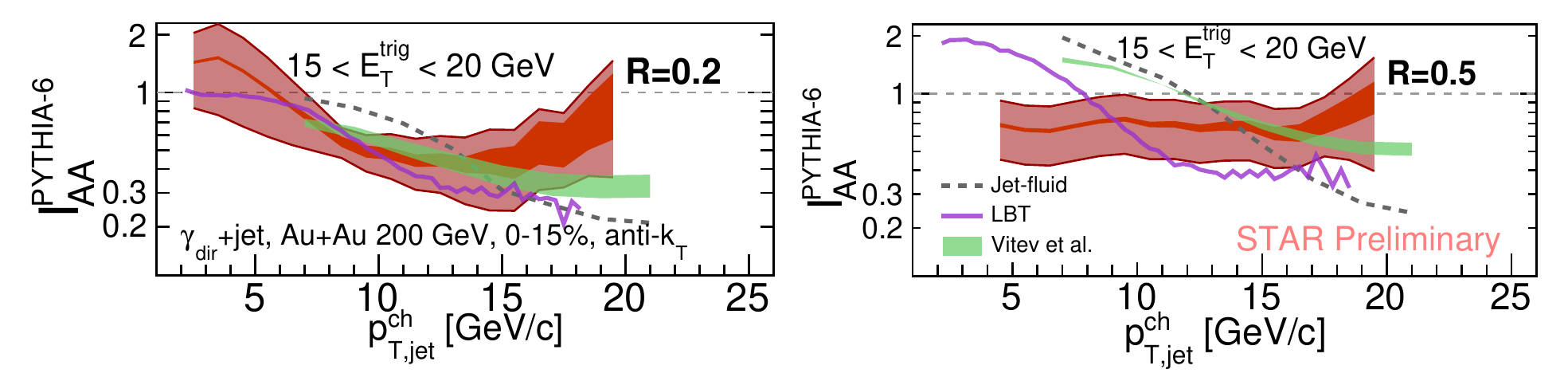}
    \caption{\gammadir+jet \IAA\ as a function of \pTjet\ for \rr\ = 0.2 and 0.5 at \sqrtsNN\ = 200 GeV\cite{Sahoo:2020kwh}.}
    \label{fig:DirPhoIAA}
\end{figure}

\begin{figure}
    \centering
    \includegraphics[width=0.9\textwidth]{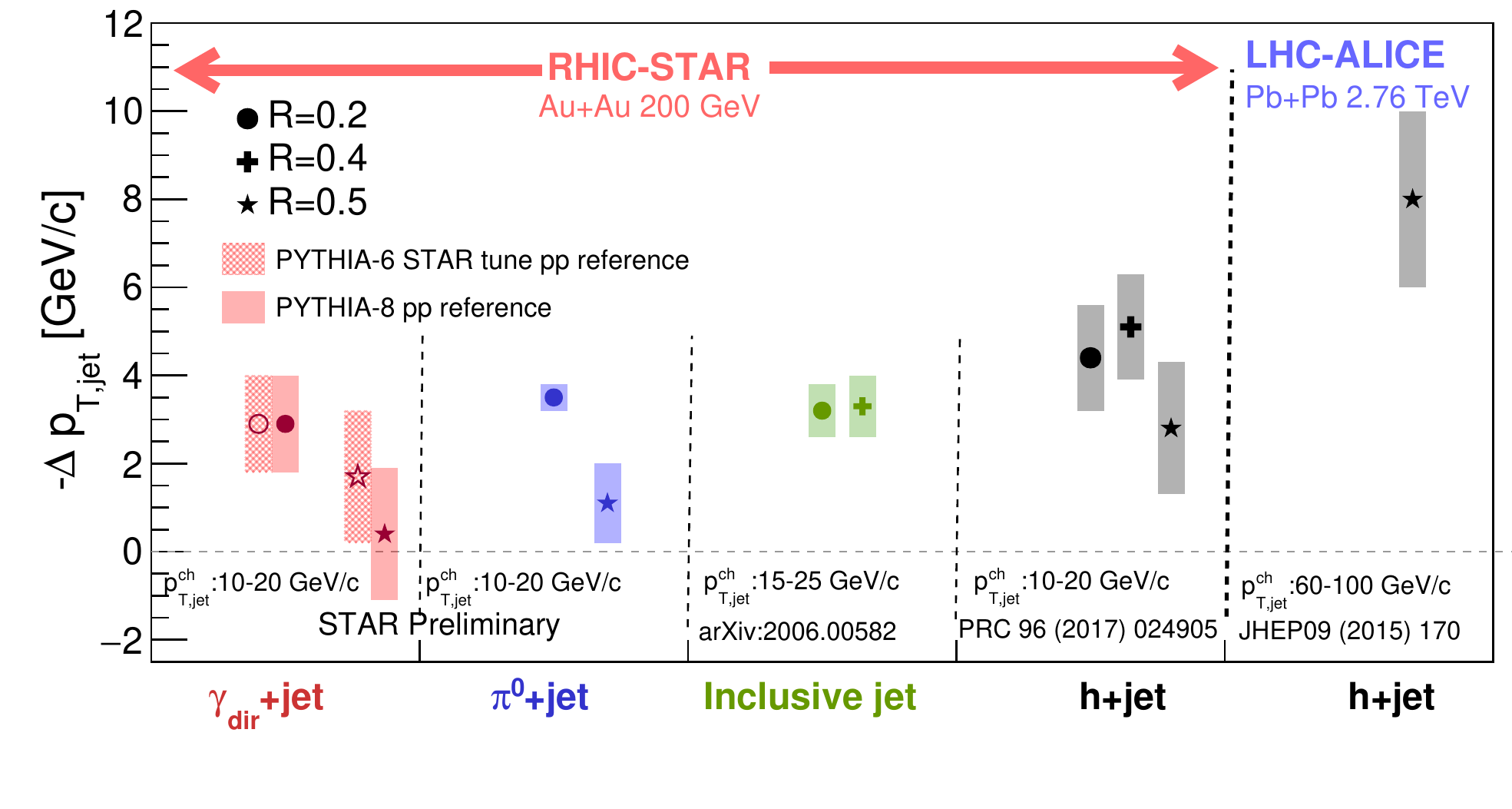}
    \caption{$-\Delta p_{\rm T,jet}$ for different observables measured at RHIC and the LHC\cite{Sahoo:2020kwh}.}
    \label{fig:PtShift}
\end{figure}

\begin{figure}
    \centering
    \includegraphics[width=1.1\textwidth]{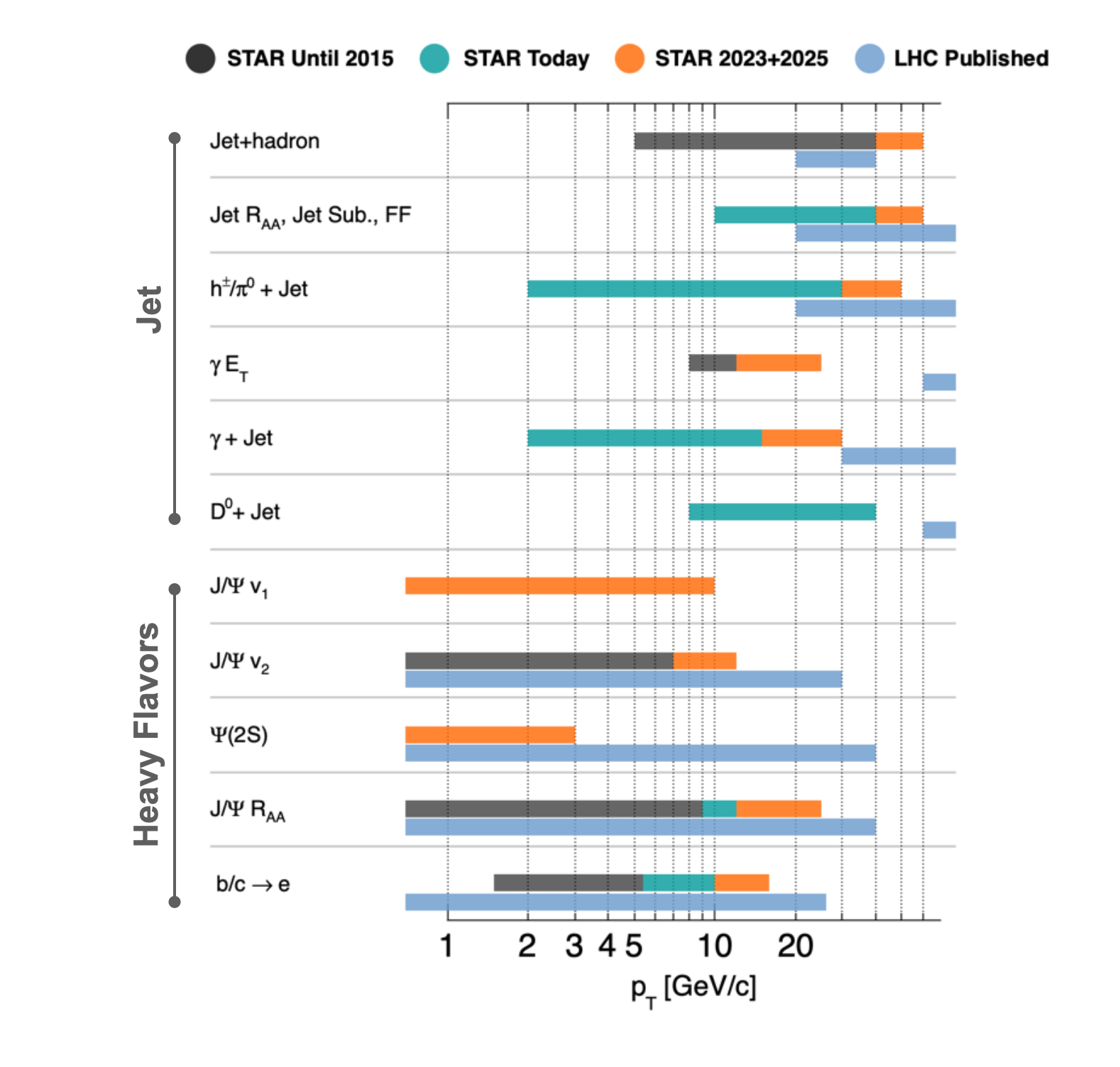}
    \caption{The kinematic coverages of the STAR hard probe measurements (past, current, and future projection) are shown and compared to the LHC (published) measurements\cite{STARBUR2225}.}
    \label{fig:STARBUR}
\end{figure}

\subsection{Semi-inclusive \gammadir+jet and hadron+jet suppression}

The STAR experiment published\cite{STAR:2017hhs} the measurement of semi-inclusive distribution of reconstructed recoil jets from a high-\pT\ trigger hadron (h+jet) in Au+Au collisions. In this measurement, the uncorrelated background contribution is mitigated by using a novel mixed-event (ME) technique. It is found that the contributions from multi-parton interactions in the recoil acceptance are negligible. \\ 

The semi-inclusive h+jet measurement enables us to perform similar \gammadir+jet and \pizero+jet measurements in STAR by combining the method used in the previous \gammadir+hadron and \pizero+hadron correlation measurements\cite{STAR:2016jdz}. In this measurement, \gammadir\ and \pizero\ are within the trigger energy ranges of 9 $<$ \ETtrig $<$ 20 GeV and 9 $<$ \ETtrig $<$ 15 GeV, respectively. However, in these proceedings only \gammadir+jet results with 15 $<$ \ETtrig $<$ 20 GeV range is discussed, which is the highest \ETtrig\ range measured in this analysis. Recoil jets are measured using anti-\kT\ algorithm with \rr\ = 0.2 and 0.5. The same ME technique as in h+jet paper is applied to subtract uncorrelated jet background. The unfolding procedure, as in h+jet paper, is applied to correct for the detector effects and heavy-ion background fluctuations. Finally, the ratio of recoil jet yield in central Au+Au collisions to that of PYTHIA-8, \IAApythia, is presented for the aforementioned jet radii. The \IAApythia\ of \gammadir+jet for \rr\ = 0.2, within 15 $<$ \ETtrig $<$ 20 GeV, shows a stronger suppression than that of \rr\ = 0.5 as shown in Fig.~\ref{fig:DirPhoIAA}, which hints at a potential \rr\ dependence of recoil jet suppression at RHIC. The upcoming results with the \pp\ data as a reference will be reported in the final publication of this measurement. Precision measurement with extended recoil jet \pTjetch\ range for \gammadir+jet is planned with the upcoming RHIC runs.

\subsection{Charged jet \pTjet\ spectrum shift: RHIC vs. LHC}
Jet suppression is commonly reported via \RAA\ and \IAA\ as a function of \pTjet. These observables convolute the effect of energy loss with the shape of the jet \pTjet\ spectrum. In order to deconvolute this, a \pTjet\ shift ($-\Delta p_{\rm T,jet}$ in Fig.~\ref{fig:PtShift}) is measured for a quantitative comparison of jet energy loss from different observables.  The STAR h+jet\cite{STAR:2017hhs}, inclusive jet, preliminary \gammadir+jet, and \pizero+jet measurements, and ALICE h+jet\cite{ALICE:2015mdb} measurements report values of $-\Delta p_{\rm T,jet}$, although within different kinematic ranges. While comparing these values as shown in Fig.~\ref{fig:PtShift}, an indication of smaller in-medium energy loss at RHIC than the LHC is seen.

\section{STAR ongoing measurements and upcoming data-taking plan}
\label{Sect:Future}
 There are several ongoing jet measurements in STAR to study the QGP medium properties in heavy-ion collisions, such as full jet reconstruction to extend the \pTjet\ reach, jet fragmentation function, jet shape, heavy-flavor jet, and recoil jet azimuthal angular correlation with trigger particles.

The STAR experiment plans to take high statistics data of Au+Au collisions at \sqrtsNN\ = 200 GeV in 2023 and 2025, and \pp\ collision data at $\sqrt{s} = 200$ GeV along with $p$+A collisions in 2024 with 28 cryo-weeks for each year\cite{STARBUR2225}. 
The kinematic coverages of various measurements related to hard probes using these datasets are presented in Fig~\ref{fig:STARBUR}.
These datasets are crucial for studying the inner-workings of QGP utilizing precision jet measurements.

\section{Summary}
\label{Sect:Summary}
 The STAR experiment has recently reported important results on jet substructure observables in \pp\ collisions, and various jet quenching observables in heavy-ion collisions to study QGP properties. Several other jet measurements are ongoing and will be presented in the near future. Besides, STAR's upcoming data-taking (during 2023-2025 RHIC runs) is crucial for the precision jet measurements with large kinematic coverages, whereas high statistics \pp\ data (2024 RHIC run) are important for providing high precision references.

\section*{Acknowledgement}
NRS is supported by the Fundamental Research Funds of Shandong University and NNSF of China:12050410235.

\bibliographystyle{ws-ijprai}

\begin{thebibliography}{0}    

\bibitem{Cunqueiro:2021wls}
L.~Cunqueiro and A.~M.~Sickles,
[arXiv:2110.14490 [nucl-ex]].


\bibitem{Salam:2010nqg} G.~P.~Salam,
Eur. Phys. J. C \textbf{67}, 637-686 (2010)
doi:10.1140/epjc/s10052-010-1314-6
[arXiv:0906.1833 [hep-ph]].

\bibitem{Larkoski:2015lea}A.~J.~Larkoski, S.~Marzani and J.~Thaler,
Phys. Rev. D \textbf{91}, no.11, 111501 (2015)
doi:10.1103/PhysRevD.91.111501
[arXiv:1502.01719 [hep-ph]].


\bibitem{STAR:2020ejj}
J. Adam \textit{et al.} [STAR],
Phys. Lett. B \textbf{811}, 135846 (2020)
doi:10.1016/j.physletb.2020.135846
[arXiv:2003.02114 [hep-ex]].

\bibitem{STAR:2021lvw}
M. Abdallah \textit{et al.} [STAR],
Phys. Rev. D \textbf{104}, no.5, 052007 (2021)
doi:10.1103/PhysRevD.104.052007
[arXiv:2103.13286 [hep-ex]].

\bibitem{STAR:2019yqm}
J. Adam \textit{et al.} [STAR],
Phys. Rev. D \textbf{100}, no.5, 052005 (2019)
doi:10.1103/PhysRevD.100.052005
[arXiv:1906.02740 [hep-ex]].

\bibitem{Gribov:1972ri}
V.~N.~Gribov and L.~N.~Lipatov,
Sov. J. Nucl. Phys. \textbf{15}, 438-450 (1972)
IPTI-381-71.

\bibitem{Dokshitzer:1977sg}
Y.~L.~Dokshitzer,
Sov. Phys. JETP \textbf{46}, 641-653 (1977)



\bibitem{Altarelli:1977zs}
G.~Altarelli and G.~Parisi,
Nucl. Phys. B \textbf{126}, 298-318 (1977)
doi:10.1016/0550-3213(77)90384-4

\bibitem{Dasgupta:2013ihk}
M.~Dasgupta, A.~Fregoso, S.~Marzani and G.~P.~Salam,
JHEP \textbf{09}, 029 (2013)
doi:10.1007/JHEP09(2013)029
[arXiv:1307.0007 [hep-ph]].

\bibitem{Larkoski:2014wba}
A.~J.~Larkoski, S.~Marzani, G.~Soyez and J.~Thaler,
JHEP \textbf{05}, 146 (2014)
doi:10.1007/JHEP05(2014)146
[arXiv:1402.2657 [hep-ph]].

\bibitem{STAR:2002ggv}
C.~Adler \textit{et al.} [STAR],
Phys. Rev. Lett. \textbf{89}, 202301 (2002)
doi:10.1103/PhysRevLett.89.202301
[arXiv:nucl-ex/0206011 [nucl-ex]].

\bibitem{STAR:2002svs}
C.~Adler \textit{et al.} [STAR],
Phys. Rev. Lett. \textbf{90}, 082302 (2003)
doi:10.1103/PhysRevLett.90.082302
[arXiv:nucl-ex/0210033 [nucl-ex]].

\bibitem{STAR:2006vcp}
J.~Adams \textit{et al.} [STAR],
Phys. Rev. Lett. \textbf{97}, 162301 (2006)
doi:10.1103/PhysRevLett.97.162301
[arXiv:nucl-ex/0604018 [nucl-ex]].

\bibitem{STAR:2020xiv}
J. Adam \textit{et al.} [STAR],
Phys. Rev. C \textbf{102}, no.5, 054913 (2020)
doi:10.1103/PhysRevC.102.054913
[arXiv:2006.00582 [nucl-ex]].

\bibitem{Vitev:2009rd}
I.~Vitev and B.~W.~Zhang,
Phys. Rev. Lett. \textbf{104}, 132001 (2010)
doi:10.1103/PhysRevLett.104.132001
[arXiv:0910.1090 [hep-ph]].

\bibitem{Chien:2015hda}
Y.~T.~Chien and I.~Vitev,
JHEP \textbf{05}, 023 (2016)
doi:10.1007/JHEP05(2016)023
[arXiv:1509.07257 [hep-ph]].

\bibitem{Casalderrey-Solana:2016jvj}
J.~Casalderrey-Solana, D.~Gulhan, G.~Milhano, D.~Pablos and K.~Rajagopal,
JHEP \textbf{03}, 135 (2017)
doi:10.1007/JHEP03(2017)135
[arXiv:1609.05842 [hep-ph]].

\bibitem{He:2015pra}
Y.~He, T.~Luo, X.~N.~Wang and Y.~Zhu,
Phys. Rev. C \textbf{91}, 054908 (2015)
[erratum: Phys. Rev. C \textbf{97}, no.1, 019902 (2018)]
doi:10.1103/PhysRevC.91.054908
[arXiv:1503.03313 [nucl-th]].

\bibitem{Ke:2018jem}
W.~Ke, Y.~Xu and S.~A.~Bass,
Phys. Rev. C \textbf{100}, no.6, 064911 (2019)
doi:10.1103/PhysRevC.100.064911
[arXiv:1810.08177 [nucl-th]].

\bibitem{STAR:2017hhs}
L. Adamczyk \textit{et al.} [STAR],
Phys. Rev. C \textbf{96}, no.2, 024905 (2017)
doi:10.1103/PhysRevC.96.024905
[arXiv:1702.01108 [nucl-ex]].

\bibitem{STAR:2016jdz}
L.~Adamczyk \textit{et al.} [STAR],
Phys. Lett. B \textbf{760}, 689-696 (2016)
doi:10.1016/j.physletb.2016.07.046
[arXiv:1604.01117 [nucl-ex]].

\bibitem{ALICE:2015mdb}
J. Adam \textit{et al.} [ALICE], 
JHEP \textbf{09}, 170 (2015)
doi:10.1007/JHEP09(2015)170
[arXiv:1506.03984 [nucl-ex]].

\bibitem{Sahoo:2020kwh}
N.~R.~Sahoo [STAR],
PoS \textbf{HardProbes2020}, 132 (2021)
doi:10.22323/1.387.0132
[arXiv:2008.08789 [nucl-ex]].

\bibitem{STARBUR2225}
STAR BUR: https://indico.bnl.gov/event/11308/


\end{thebibliography}

\end{document}